# 3D high resolution seismic model with depth: A relevant guide for Andra deep geological repository project

J.L. Mari [1], B. Yven [2]

1. IFP Énergies nouvelles, IFP school, 92852, Rueil Malmaison Cedex, France, jean-luc.mari@ifpen.fr
2. Andra, 1-7 parc de la Croix Blanche, 92298 Châtenay-Malabry Cedex, France, beatrice.yven@andra.fr

## Abstract

In the context of deep geological disposal of high level radioactive wastes, the French National Radioactive Waste Management Agency (Andra) has conducted an extensive characterization of the Callovo-Oxfordian argillaceous rock and surrounding formations in the Eastern Paris Basin. For such characterization, a detailed 3D geological model is needed. This paper shows the procedure used for building the 3D model with depth by combining time-to-depth conversion of seismic horizons, consistent seismic velocity model and elastic impedance in time. It also shows how the 3D model is used for mechanical and hydrogeological studies. The 3D field data example illustrates the potential of the proposed depth conversion procedure for estimating a density model with depth. The geological model shows good agreement with logging data obtained at a reference well, a closest borehole located in the vicinity of the 3D seismic area.

Modeling of mechanical parameters such as shear modulus, young modulus, bulk modulus indicates weak variability of these parameters which confirm the homogeneity of the Callovo-Oxfordian clay. 3D modeling of a permeability index (Ik-Seis) computed from seismic attributes (instantaneous frequency, envelope, elastic impedance) and validated at the reference borehole shows promising potential for supporting hydrogeological simulation.

# Key words

# 1. Introduction

3D seismic blocks and logging data, mainly acoustic and density logs, are often used for geological model building with time data. The geological model must be then converted from time data to depth data. A geostatistical approach for time-to-depth conversion of seismic horizons is often used in many geomodeling projects. The time-to-depth conversion of the selected seismic horizons by Bayesian kriging (Sandjivy and Shtuka, 2009) can be used to compute a time-to-depth conversion model at the time sampling rate (1 ms). The 3D depth conversion model and the 3D impedance block allow the computation of interval velocity and density blocks. It is also possible to extract specific attributes from seismic sections to compute a seismic index (Ik-Seis factor) leading to a better understanding of the distribution of the porous and permeable bodies (Mari and Guillemot, 2012). The Ik-Seis, density and velocity sections can be fruitfully combined to do a mechanical and hydrogeological study with depth.

The proposed procedure has been applied both on 2D and 3D data sets. The seismic data come from High Resolution seismic surveys conducted in the vicinity of the French National Radioactive Waste Management Agency (Andra) Meuse/Haute-Marne Underground Research Laboratory (URL) located in the Meuse department in Eastern France. .

Within its scientific and experimental programs devoted to investigate the disposal of high level radioactive wastes in a Callovo-Oxfordian argillaceous rock (Cox) of the eastern Paris Basin, Andra has conducted an extensive characterization of the Cox and the Oxfordian and Dogger limestones directly located above and below the target formation. More than 50 deep wells were dedicated to the geological, hydrogeological and geomechanical characterization of the Jurassic and Triassic geological series. In addition, 2D and 3D seismic data have been recorded to build a 3D structural and stratigraphic model and to estimate the variability of the hydrogeological and geomechanical parameters of the geological formations. The main purpose of this study is to describe the results of the 3D modeling of the geometry and properties of the Cox host rock in which a repository may be built. These results will serve as the basis for further simulations to assess the repository.

After a short review of the geological setting, we describe the seismic spreads which has been used to record high resolution seismic data, the processing sequence applied to the seismic data set and the procedure used to perform the time to depth conversion of the time migrated seismic sections. Finally, we show how the density seismic sections are used to detect shaly layers and how the Ik-Seis attribute can be a qualitative indicator used to detect the presence of water-bearing layers. The results obtained are compared with logging data recorded in wells EST431 and EST433 located on the 2D seismic line 07EST10.

## 2. Geological setting

The region where Andra URL is located forms a geologically simple domain of the Paris basin, with a succession of marine limestone, marl and clay rock layers. The dip of the layers is low, around 1° to 1.5° towards the north-west. The host formation, a clay rock 155 million years old, at least 130 meters thick and located between 400 and 600 meters below ground, is referred to as the Callovo-Oxfordian argillites (Cox). The top of the formation has a greater carbonate content and includes interbedded clayey layers and carbonate rock. The Cox is more homogeneous in its central part with a clay-mineral concentration of 45-50%, which corresponds to a maximum of flooding within the depositional environment of the area. A detailed study of the spatial variability of Cox geological and physical properties is found in Garcia et al. (2011). This formation is embedded by two thick limestone formations of Dogger and Oxfordian age. The Dogger (Bathonian, Bajocian) corresponds to the development of carbonate platforms in the Paris Basin (Gaumet, 1997; Purser, 1980). The clayey formation is separated from the Dogger limestone by a regional paleo-erosion surface. The Bathonian limestone formation is the first aquifer encountered in the Jurassic series.

Above the Cox, an Oxfordian (Middle and Late) carbonate platform developed with reefs and oobioclastic facies. The Oxfordian limestone formation represents the second Jurassic aquifer. In the area investigated by seismic surveys, the outcropping formations are Kimmeridgian marls and Kimmeridgian and Tithonian limestones with a thin deposit of Cretaceous.

The readers should refer to Brigaud et al. (this issue) for a detailed description of the stratigraphic architecture of the Jurassic carbonate platforms. Figure 1 shows the geological description of the Jurassic and Triassic series at well EST433 and the location of the interpreted seismic horizons.

The structural deformations have remained small for the past 150 million years, as in the rest of the Paris basin (Guillocheau et al., 2000; Megnien, 1980). They are essentially limited to the Gondrecourt and Marne grabens, on the boundary of the sector studied (André et al., 2004; Rocher et al., 2004). The geological and geophysical studies have shown that the Callovo-Oxfordian layer is regular and practically flat between these faults. Available data also confirm that the region has very low seismicity.

## 3. Seismic acquisition and processing

Figure 2 shows the 3D seismic data acquisition area and the location of the selected seismic lines in the transposition area: the 2D line 07EST10 and the in line 405 extracted from the 3D block. The 2D line 07EST10 is oriented W-E and the in line 405 is oriented NNW-SSE. No wells being located in the 3D area, three additional 2D lines have been recorded to calibrate the 3D data on 3 wells situated outside of the 3D area. Well EST433 is located on one of the additional 2D line 07EST10 in the vicinity of the CMP 654 (Common Mid Point).

The 3D design is a cross spread. The active spread is composed of 12 receiver lines with 120 stations each. The source lines are perpendicular to the receiver lines. The receiver and source line spacing's are respectively 80 m and 120 m. The receiver and source point spacing's are 20 m. The source is a vibroseis source generating a signal in the 14 - 140 Hz frequency bandwidth. The bin size is 10 x 10 m. The nominal fold is 60. The in line 405 is composed of 544 CMP points.

The 2D design is a split dip spread, composed of 120 stations. The receiver and source point spacing's are 25 m. The source is a vibroseis source generating a signal in the 14 - 140 Hz frequency bandwidth. The bin size is 12.5 m. The nominal fold is 60. The selected part of the 2D line 07EST10 is composed of 727 CMP points.

The same processing sequence has been applied both on the 2D and 3D data sets. It is a conventional seismic sequence which includes:

- data editing,
- minimum phase conversion,
- amplitude recovery (using $T^2$ law),
- surface consistent amplitude compensation (source and receiver),
- de-noise and wave separation on shots,
- statics application (data shifted to floating datum),
- surface consistent deconvolution,
- velocity analysis (on 480 m grid),

- surface consistent residual statics,
- second pass velocity analysis (on 240 m grid),
- surface consistent amplitude compensation (source, receiver and offset),
- interpolation and regularization in offset planes and noise attenuation,
- velocity model updating (residual move out on 240m grid),
- QC : 60 fold CMP stack (0-1400 m offsets ) with static to the final datum (450 m MSL),
- pre-stack time migration (time shifted to the final datum: 450 m MSL),
- Q compensation,
- Noise attenuation and phase conversion (statistical to zero phase),
- Acquisition foot print attenuation and band pass filter (15, 20 - 140, 160 Hz).

Andra supplied a geological model (distribution of velocities in depth) to compute static corrections. An up-hole survey (20 wells) was conducted to calibrate the model of static corrections.

Figure 3 shows the 2 selected seismic lines (2D line 07EST10 and the in line 405) after the processing sequence just described. For all the seismic sections and logs, the depth reference is the seismic datum plane (450 m MSL). The selected lines will be used to illustrate the different steps of the procedure and the results obtained.

During the migration process (Robein, 2003), the data are migrated and sorted in common image gathers (CIG) to update the velocity model before stacking. Each image gather is composed of migrated seismic traces which are functions of time (time migration) and offset or angle. The offset-angle conversion can be done during the migration process or after by using the velocity model. The common image gather after offset-angle conversion are used to perform Amplitude versus Angle analysis (Castagna, 1993; Walden, 1991) and elastic inversion (EI). The amplitudes of small angle (near-offset or intercept) migrated stacks relate to change in acoustic impedance (AI) and can be inverted back to AI using post-stack inversion algorithm (acoustic inversion). The acoustic impedance is a simple function of P-wave velocity Vp and density ρ (AI = ρ.Vp). The elastic impedance (EI) is a function of P-wave velocity Vp, S-wave velocity Vs, density ρ and incidence angle (Connolly, 1999). The amplitudes of angle migrated stacks can be inverted back to impedances Ip (Ip = AI = ρ.Vp) and Is (Is = ρ.Vs) using a linearization of Zoeppritz equations for P-wave reflectivity (Shuey, 1985). Such a processing refers to elastic inversion (Whitcombe et al., 2002). A model-based elastic inversion (a priori impedance model obtained from well data), applied to the angle migrated stacks, provides 3D impedance blocks (Ip and Is blocks). The acoustic impedance sections (Ip-sections) are shown in Figure 4 for the selected seismic profiles.

A geostatistical approach for time-to-depth conversion of seismic horizons is often used in many geo-modelling projects. From a geostatistical point of view, the time-to-depth conversion of seismic horizons is a classical estimation problem involving one or more secondary variables. The converted depth and associated uncertainty can be estimated using a kriging method which can be constrained by the well markers, velocity model and interpreted horizons. For the multilayer case, the kriging estimator should take into account all the relationships between horizons determined by the velocity model associated to each layer.

The more appropriate kriging method for this problem is the Bayesian Kriging (BK) (Sandjivy and Shtuka, 2009; Omre and Halvorsen, 1989; Omre, 1987). Bayesian approach provides an excellent estimator which is more general than the traditional kriging with external drift(s) and fits very well to the needs for time-to-depth conversion of seismic horizons. The advantage of BK compared to other estimators consists in the fact that the uncertainty on the trend velocity model and the local uncertainty defined by the uncertainty of interpreted time maps and local fluctuations of interval velocities can be managed simultaneously.

The input information for BK are:
- Two-way-time (TWT) maps for interpreted horizons,
- Well markers for each horizon,
- Prior velocity model and associated uncertainty for each layer,
- Local uncertainty definition for each time map (picking uncertainty, and spatial variogram definition),
- Local uncertainty definition of interval velocity for each layer (local velocity fluctuations around the trend model, and spatial variogram definition).

As any Kriging based estimator, the BK provides:

- The estimated variable (estimated depth for each horizon),
- Variance of estimation (associated uncertainty of estimated depth).

The use of BK in depth conversion has the advantage to combine the prior knowledge of the velocity model with a certain degree of uncertainty and the well data. All sources of uncertainty (velocity and time) are integrated in a consistent way in a unique probabilistic model used for estimation or simulation.

For each selected horizon, the BK provides its estimated depth Z associated with its time t. The " Z versus t " data set is interpolated in the whole space (3D block) at the time sampling rate (1 ms) in order to obtain a time to depth conversion model, using the impedance sections to estimate the short wavelength variations of the velocity model.

The time to depth conversion procedure is illustrated via the line 07EST10 (see location map, Figure 2). 10 seismic horizons numbered from 1 to 10 have been picked in time and depth converted. The 10 seismic horizons are:

1. Top of Kimmeridgian White Limestone
2. Top of Porous Horizon HP4
3. Top of Lower Oxfordian (Top of target interval)
4. Top of Upper Callovian (RIO)
5. Top of Carbonated Dogger (Base of target interval)
6. Base of Argillaceous limestone and marls
7. Base of Carbonated Dogger
8. Top of Domerian
9. Base of Lias (base of Gryphees limestone)
10. Top of Beaumont dolomite.

Figure 5 shows the selected times of the 10 seismic horizons, the time-to-depth conversion model, the depth conversion of the 10 horizons and the results of the depth conversion for the seismic trace located at the CMP 654 located near the well EST433. Figure 6 shows the two amplitude seismic sections in depth.

The time-to-depth conversion model can be used to convert in depth any type of seismic sections (amplitude, velocity, density, shear modulus, Ik-Seis factor ...).

## 4. Structural study of the Callovo-Oxfordian formation

Figure 7 shows the elevation map below sea level of the base of the Cox formation obtained directly from the seismic data interpretation and the time-to-depth conversion process (horizon 5). The isolines are gently dipping towards the north-west (center of Paris Basin) and are subparallel. This dip has been computed and shows an average value of 1°. The difference of elevation from the east to the west in the ZIRA is 110 m. No evidence of fault may be interpreted from this map. The same conclusions arise from the maps directly below and above the Cox formation.

Figure 8 shows the thickness of the Cox formation. The thickness increases from the south-west (142 m) to the north-east (162 m) in the ZIRA with a mean thickness of 153 m. The direction of thickening is coherent with the main source and direction of sediments transport.

## 5. Geomechanical study

The 3D depth conversion model is used for computing of an interval P-wave velocity block (at a thin sampling rate 1 m) which is compared with the acoustic impedance block to estimate a density block. Figure 9 shows the P-wave velocity and density distributions in depth. The low values of density and velocity in the 600-750 m depth interval for the line IL-405 and in the 500-700 m depth interval for the line 07EST10 are associated with the target interval situated between the top of Lower Oxfordian and the top of the Carbonated Dogger. The knowledge of the density ($\rho$) and velocity ($V_p$, $V_s$) distributions allows the computation of mechanical modules such as shear modulus, Young modulus, bulk modulus and Poisson's ratio. Figure 10 shows the S-wave velocity and shear modulus distributions in depth. The seismic data obtained at CMP 654 have been compared with logging data recorded at well EST433.

The comparison results are shown in Figures 11 and 12. The correlation coefficients between the 2 datasets are high (larger than 0.9) for the velocities ($V_p$, $V_s$) and the shear modulus (Mu) and low (smaller than 0.6) for the density ($\rho$). Furthermore, the seismic densities are underestimated in comparison to the logging densities (Figure 12). The sections of mechanical parameters ($V_p$, $V_s$, $\rho$, shear modulus (Mu), Young modulus and bulk modulus) have been used to study the variability of the mechanical properties along the two seismic profiles. The results are given in table 1. The weak variability of the mechanical modules confirms the homogeneity of the target formation. The density sections (Figure 9, top) have been flattened at the top of the White Kimmeridgian limestone and the histograms of the density values have been computed for different depth intervals.

Figure 13 (top) shows the following results:

- Between the top of Lower Oxfordian and the top of carbonated Dogger: It is the depth interval of the Cox formation. On the histogram, one single mode is observed. The average density is 2.34 g/cm$^3$ and the associated standard deviation is 0.03 g/cm$^3$.
- Between the top and the base of carbonated Dogger: One single mode is observed on the histogram. The average density is 2.56 g/cm$^3$ and the associated standard deviation is 0.04 g/cm$^3$.
- Between the top of White Kimmeridgian limestone and the top of Beaumont dolomites (over a depth thickness of 1200 m), three modes can be observed on the histogram.

In the Callovo-Oxfordian, the highest value of the density is 2.38 g/cm$^3$. A threshold of 2.38 g/cm$^3$ has been applied to the density section to compute a shale indicator which equals 0 if the density value is smaller than 2.38 g/cm$^3$ and 1 if not. The shale indicator sections are shown in Figure 13 (bottom). The shale indicator obtained at CMP 654 has been compared with a good agreement with the indicator in the lithological log of well EST433 (Figure 1). This result allows using the shale indicator sections to study the spatial lithological variability of the shale layers from the Triassic to the Upper Jurassic formations.

|  | Average Value | Standard deviation |
|---|---|---|
| **P-wave velocity (m/s)** | 3000 | 210 |
| **S-wave velocity (m/s)** | 1700 | 200 |
| **Density (g/cm$^3$)** | 2.34 | 0.03 |
| **Poisson's ratio** | 0.28 | 0.06 |
| **Bulk modulus (GPa)** | 13 | 2.7 |
| **Shear modulus (GPa)** | 6.5 | 1.5 |
| **Young modulus (GPa)** | 17 | 3.1 |

Table 1: Variability of mechanical parameters in the Callovo-Oxfordian formation

## 6. Hydrogeological study

The impedance model (Ip block) can be converted into porosity by using an empirical relationship between porosity and acoustic impedance established at well locations. To model porosities, another option is to use porosity at wells location and interpolate between the wells by means of kriging. Partly due to the small number of wells, this outcome is really smooth and usually does seem geologically consistent. More dense information can be integrated in order to improve the estimation of porosity. As porosity is linked to acoustic impedance, it is relevant to use dense seismic acoustic impedance information. So a collocated cokriging of porosity integrating the seismic information is performed by using the normalized acoustic impedance as the secondary variable (Bourges et al., 2012). 3D cube makes it possible to provide 3D imaging of the connectivity of the porous bodies (Mari and Delay, 2011). Core analysis is usually carried out to establish porosity vs. permeability laws (Zinszner and Pellerin, 2007). It has been shown that it is possible to extract new attributes from seismic sections, leading to a better understanding of the distribution of the porous and permeable bodies (Mari and Guillemot, 2012). The attributes are also used to detect the impermeable layers.

Laboratory experiments (Morlier and Sarda, 1971) have shown that the acoustic attenuation of a clean formation can be expressed in terms of three structural parameters: porosity, permeability and specific surface. Both theoretical and experimental studies have identified the relation between acoustic attenuation and petrophysical parameters:

$$\delta = (C.S/\varphi)(2\pi.k.f.\rho_f/\mu)^{1/3} \qquad (1)$$

With δ attenuation (dB/cm), f frequency (Hz), ρf fluid density, µ fluid viscosity (centipoise) φ porosity, S Specific surface (cm$^2$/cm$^3$), C calibration coefficient and k permeability (mD).

Fabricius et al. (2007) have found that the specific surface with respect to grain volume (Sg) is apparently independent from porosity. In an attempt to remove the porosity effect on Vp/Vs and mimic a reflected φ vs log (Sg) trend, they propose to use the following relationship between porosity φ, Vp/Vs and Sg:

$$\log(Sg.m) = a.\varphi + b.(Vp/Vs) + c \quad \text{with} \quad Sg = S/(1-\varphi) \qquad (2)$$

where it should be observed that Sg is multiplied by m to make Sg dimensionless. To establish eq 2, Fabricius at al. (2007) have looked at ultrasonic data, porosity, and permeability of 114 carbonate core plugs.

In practice, the parameter Ik-Seis (Indicator (**I**) of permeability (**k**) from acoustic or seismic (**Seis**) data) computed from equation 1 is proportional to the permeability k.

$$\text{Ik-Seis} = (\varphi.\delta/S)^3/f = (\varphi/SQ)^3/f \qquad (3)$$

with f P-wave frequency, Q quality factor, δ attenuation, S specific surface and φ porosity.

The formation permeability indicator Ik-Seis can be obtained via the computation of 4 input data: P-wave frequency and attenuation, porosity and specific surface. The procedure has been firstly conducted in acoustic logging to estimate permeability of porous layers and to detect water inflows (Mari and Guillemot, 2012). In seismic, the processing is performed to measure these parameters. The analytic signal is used to compute the instantaneous frequency and attenuation (Q factor). The porosity and specific surface are computed from seismic impedances obtained by elastic inversion of the migrated seismic sections. In the domain of seismic frequencies, the Ik-Seis factor can only be seen as a relative indicator which varies from 0 for less porous and permeable bodies to 1 for more porous and permeable bodies. More information concerning the data processing and analysis is given in Mari and Guillemot (2012).

Well EST 433 was used to study formations ranging from Oxfordian to Trias. Well EST 431 was drilled to complement the geological and hydrogeological knowledge of the Oxfordian formation. This formation consists essentially of limestone deposited in a vast sedimentary platform. The limestone facies, which vary from one borehole to another, are generally bio-detritic with reef constructions. In this formation, porosity ranges between 5 and 20% and "porous horizons" of kilometric extension have been identified. The observed water inflows are usually located in high porosity zones. During the drilling, water inflows were detected at - 456 m and - 503 m (reference depth: seismic datum plane). The acoustic tool used for the field experiment is a flexible monopole tool with two pairs of receivers: a pair of near receivers (1 and 1.25 m offsets) and a pair of far receivers (3 m (R1) and 3.25 m (R2) offsets). The data have been recorded through the far offset configuration. The sampling depth interval is 10 cm. The sampling time interval is 5 microseconds. The length of recording is 5 ms. The acoustic log has been run in the Oxfordian carbonate formation, in the 420-600 m depth interval. Figure 14 (top left) shows the R1 and R2 constant offset sections in the 0.5-1.25 ms time interval where the refracted P-waves appear. The processing of the acoustic data has been described in detail by Mari et al. (2011). Figure 14 (top right and bottom left) is a display of acoustic logs: P-wave velocity ($V_P$), P-wave frequency, acoustic porosity, P-wave attenuation. For a clean formation, if the matrix and fluid velocities are known, an acoustic porosity log can be computed from the acoustic $V_p$ velocities using the formula given by Wyllie et al. (1956) expressed in velocities. The matrix velocity value has been chosen at 6300 m/s, and the fluid velocity at 1500 m/s (the formation fluid is water). The acoustic porosity log, valid only in the clean part of the formation, shows a strong correlation (correlation coefficient: 0.86) with a NMR porosity log (not displayed here) recorded in the well. Figure 14 (bottom right) shows a comparison between acoustic velocities at wells EST431 (black curve) and EST433 (red curve) and between acoustic velocities at well EST431 (black curve) and seismic velocities at CMP 654 (red curve). The correlation coefficients are high (0.88).

The attenuation (expressed in dB/m) of the formation is computed from the first eigensection (obtained by Singular Value Decomposition SVD) of the refracted P-wave acoustic signal recorded by the two adjacent receivers of the acoustic tool. The results obtained by the SVD processing procedure are shown in Figure 15 (top). The Ik-Seis log detects three permeable zones at 456 m, between 490 and 530 m, and at 594 m. The permeable zone located at 594 m corresponds to a high value of conductivity and is characterized by a low porosity (6 %), a 10 dB/m attenuation, but a significant decrease of the P-wave frequency. The hydraulic tests and conductivity measurements conducted later on did not confirm the inflow at 456 m seen during the drilling, but have validated the 490-530 m and 594 m permeable zones detected by the acoustic logging.

The analysis of the acoustic waves recorded simultaneously on both receivers of the acoustic tool is used to compute additional logs defined as acoustic attributes useful for the characterization of the formation, such as amplitude, shape index and attenuation logs. The results obtained are optimum if the studied wave is extracted from the records and if the signal to noise ratio is high. We show the benefit of using Singular Value Decomposition (SVD) for that purpose (Glangeaud and Mari, 2000). The SVD processing is done on the 2 constant offset sections independently, in a 5 traces (N=5) depth running window. After flattening of each constant offset section with the picked times of the refracted wave, the refracted wave signal space is given by the first eigensection obtained by SVD:

$$\underline{r}^{sig} = \lambda_1 \underline{u}_1 \underline{v}_1^T \qquad (4)$$

$\underline{v}_1$ is the first singular vector giving the time dependence, hence named normalized wavelet, $\underline{u}_1$ is the first singular vector giving the amplitude in depth, therefore called propagation vector and $\lambda_1$ the associated eigenvalue. The amplitude variation of the refracted wavelet over the depth interval is $\lambda_1 \underline{u}_1$. Figure 15 (top) shows the normalized wavelet ($\underline{v}_1$) and the associated amplitude ($\lambda_1 \underline{u}_1$) log versus depth, for the two constant offset sections associated with the two receivers (R1 and R2) of the acoustic tool. The amplitude logs have been used to compute the attenuation log (Figure 14, bottom left) expressed in dB/m. The correlation coefficient (Figure 15, bottom left) between the two normalized wavelets has been computed at each depth. We can notice some anomalies at local depth (446, 465, 550, 580 and 590 m) and a significant decrease of the correlation coefficient in the 490-530 m depth interval. The interval corresponds to the porous and permeable zone detected by the Ik-Seis factor (Figure 15, bottom right). It is therefore suggested that changes in phase or distortion of the acoustic signal is linked to propagation through a porous and permeable zone. The distortions can be measured by a shape index attribute. To measure the shape variation, an acoustic attribute, named Ic, independent of the energy of the source, has been introduced (Lebreton and Morlier, 1986). The Ic parameter is given by the following equation:

$$Ic = ((A_2 + A_3)/A_1)^n \qquad (5)$$

where $A_1$, $A_2$ and $A_3$ are the amplitudes of the first three arches, respectively, of the studied signal and n an exponent.

The shape index is computed from formula 5 with an exponent value of 3 (n = 3). Figure 15 (top) allows the comparison between the two shape index logs (Ic-R1 and Ic-R2). The shape index logs highlight anomalic zones, in the 440 – 470 m depth interval with a maximum value at 454 m, and in the 490 – 530 m depth interval with a peak at 495 m which corresponds to the porous layer HP4. In order to reduce the noise to extract the common component of the two shape index logs (Ic-R1 and Ic-R2), the geometric mean of the two has been computed. The resulting shape index log Ic is compared with the Ik-Seis log and the correlation coefficient log (Figure 15, bottom left). The comparison shows a good coherence between the Ic log and the correlation coefficient log. The permeable and porous layers in the 490 – 530 m depth interval are seen both by the shape index log and by the Ik-Seis log. Figure 15 (bottom left) shows a comparison between Ik-Seis factors computed from acoustic data (well EST431, red curve) and seismic data (CMP 654, black curve). The seismic Ik-Seis factor is the envelope of the acoustic Ik-Seis factor, due to the different of vertical resolution between the seismic and acoustic data. The seismic Ik-Seis factor points out to permeable zones in the 440 – 470 m and 490 – 530 m depth intervals. The flow at 594 m cannot be detected due to the lack of resolution. The vertical resolution is proportional to the seismic wavelength; a quarter of the wavelength is usually adopted. The seismic velocity blocks (Figure 9) and the instantaneous frequencies have been used to compute the vertical resolution seismic sections (Figure 16, top). The Ik-seis seismic sections are shown in Figure 16 (bottom). The shaly layers, associated with a shale index equal to 0 (Figure 13) have the lowest Ik-Seis value. The Ik-Seis sections show that the Callovo-Oxfordian is a very low permeable formation with a low variability of the Ik-Seis factor (average value: 0.06, standard deviation: 0.1). Porous and permeable bodies in the Oxfordian have been detected and confirmed by acoustic logging. In the two Ik-Seis seismic sections, the Oxfordian limestone appears to have a higher Ik-Seis factor than the Dogger limestone. In the Dogger, the Ik-Seis factor is generally low except directly above the base of the target interval. However this level of higher Ik-Seis seems to vary spatially. Figure 17 is a comparative display of Ik-Seis factor at CMP 654, free fluid NMR porosity, gamma-ray and sonic logs at the well EST433. A good correlation between the Ik-Seis factor, the NMR porosity and the location of the porous levels (Hp1-Hp4 in the Oxfordian, HpD1 and HpD2 in the Dogger) can be noted. This result allows to perform a detailed study the 3D spatial variability of the Ik-Seis factor in the two carbonate formation.

## 7. Conclusions

We have described the workflow which was used to build a geological model with depth combining time-to-depth conversion of seismic horizons, consistent seismic velocity model and elastic impedance inversion in time. The field data example illustrates the potential of the proposed depth conversion procedure for estimating a density model in depth. The knowledge of the density distribution allows the computation of mechanical parameters such as shear modulus, Young modulus or bulk modulus. The seismic results obtained in the vicinity of well EST433 have been compared with the information obtained by logging (mainly velocity and density measurements). The correlation coefficients between the seismic and logging data are high (larger than 0.9) for the velocities. If the results (time depth conversion, velocities, density, mechanical parameters) remain in the expected range of uncertainties, the proposed workflow could be used to build a reliable 3D model with depth from the full seismic 3D block.

On the studied seismic lines (07EST10 and IL-405), the variability of the seismic parameters (P-wave and S-wave velocities, density, shear modulus, Young modulus and bulk modulus) is weak in the Callovo-Oxfordian leading us to conclude that it is an homogeneous formation.

The Ik-Seis sections in depth have shown that the Callovo-Oxfordian is a very low permeable formation and have detected porous and permeable bodies in the Oxfordian and at the top of the Dogger limestone. The results have been confirmed at well (EST431) using full waveform acoustic data.

A permeability index in 3D (Ik-Seis block) has a promising potential for supporting hydrogeological assessment such as determining and mapping in 3D the porosity and permeability variability of certain geological formations and hydrogeological modeling.

## 8. Acknowledgements

This work was financed by Andra. We thank Michel Hayet and Philippe Dubreuilh for their comments and suggestions. The authors acknowledge Arben Shtuka (Seisquare) for very useful discussions on various occasions, specifically for his experience in the use of geostatistical methods for time to depth conversion.

Figure 1: EST 433 geological log and seismic horizons. 1 Top of Kimmeridgian White Limestones, 2 Top of Porous Horizon HP4, 3 Top of Lower Oxfordian (Top of target interval), 4 Top of Upper Callovian (RIO), 5 Top of Carbonated Dogger (Base of target interval), 6 Base of Argillaceous limestones and marls, 7 Base of Carbonated Dogger, 8 Top of Domerian, 9 Base of Lias (base of Gryphees limestone, 10 Top of Beaumont dolomite.

Figure 2: Location map: 3D seismic area, seismic lines (IL 405 and 07EST10) and Well EST433. The 2D line 07EST10 is oriented W-E and the in line 405 is oriented NNW-SSE.

Figure 3: Amplitude seismic sections in time.

Figure 4: Acoustic impedance (g/cm$^3$ m/s) sections in time.

Figure 5: Time-to-depth conversion by Bayesian Kriging.

Figure 6: Amplitude seismic sections in depth.

Figure 7: Elevation map of the base of the Callovo-Oxfordian formation.

Figure 8: Thickness map of the Callovo-Oxfordian formation.

Figure 9: P-wave velocity and density seismic sections in depth.

Figure 10: S-wave velocity and shear modulus seismic sections in depth.

Figure 11: Comparison between seismic and logging data at well EST433 (CMP 654, line 07EST10).

Figure 12: Cross plots between seismic (CMP 654, line 07EST10) and logging data at well EST433.

Figure 13: Density analysis and shale indicator sections in depth

Figure 14: Acoustic logging at well EST431. Top left: acoustic sections recorded on receivers R1 and R2 of the acoustic logging tool. Top right: P-wave velocity and frequency logs. Bottom left: porosity and attenuation logs. Bottom right: comparison between acoustic and seismic velocities.

Figure 15: Permeability estimation from acoustic logs at well EST431. Top: Analysis of refracted waves by SVD (Wavelet, amplitude, shape index) on receivers R1 and R2 of the acoustic logging tool. Bottom left: predicted permeability (Ik-Seis), shape index and correlation coefficient logs. Bottom right: comparison between Ik-Seis factors computed from acoustic data (well EST431) and seismic data (CMP 654).

Figure 16: Vertical resolution and Ik-Seis seismic sections in depth.

Figure 17: Correlation between permeability index Ik-Seis and logging data (NMR porosity, Gamma-ray and sonic logs).

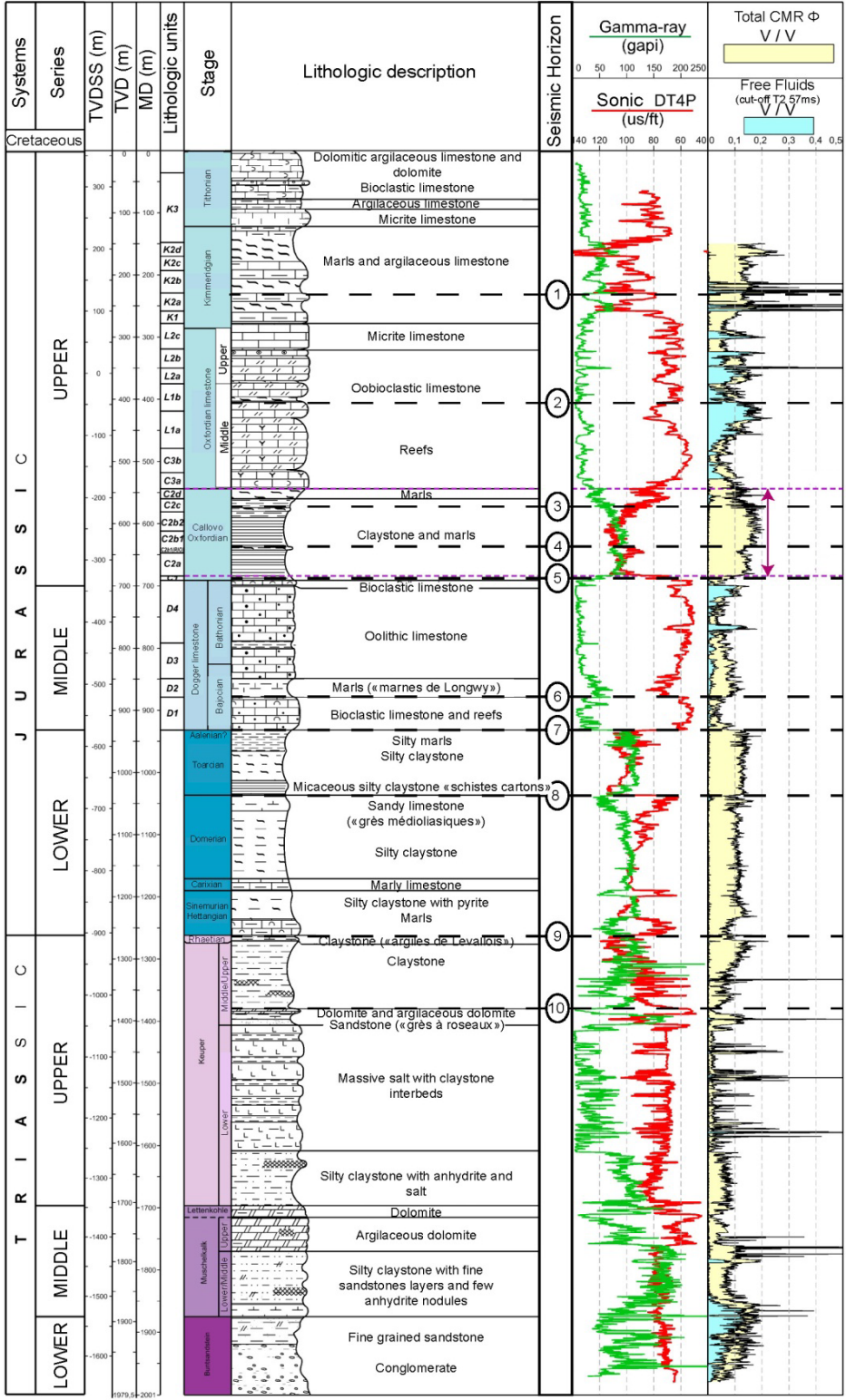

Figure 1

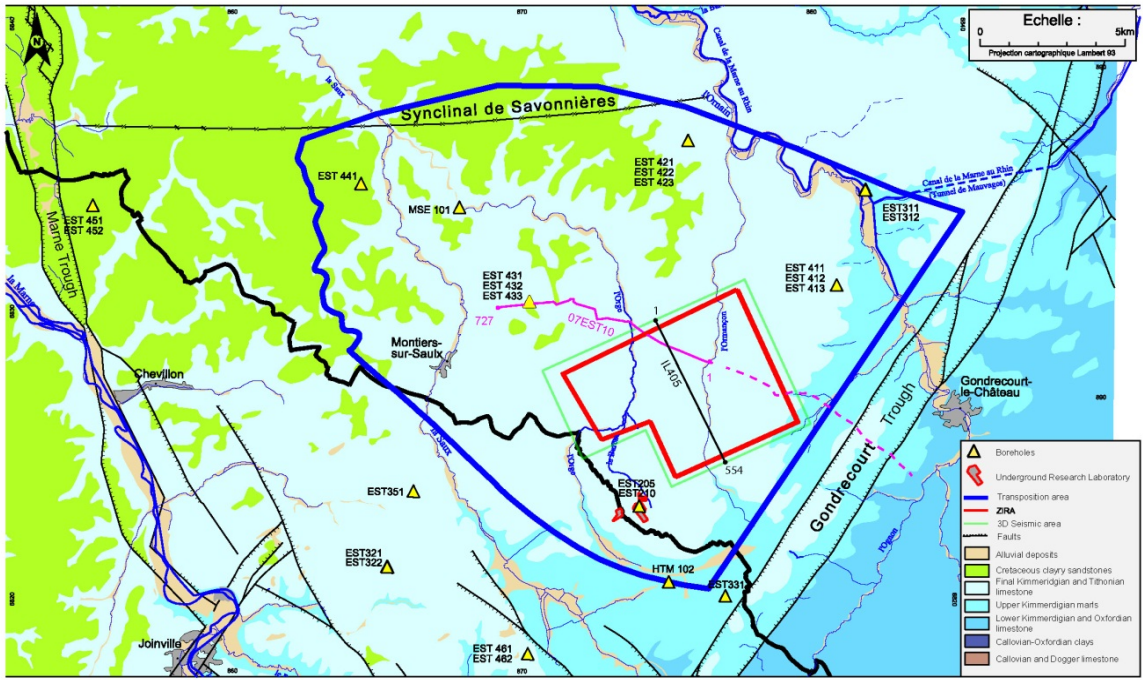

Figure 2

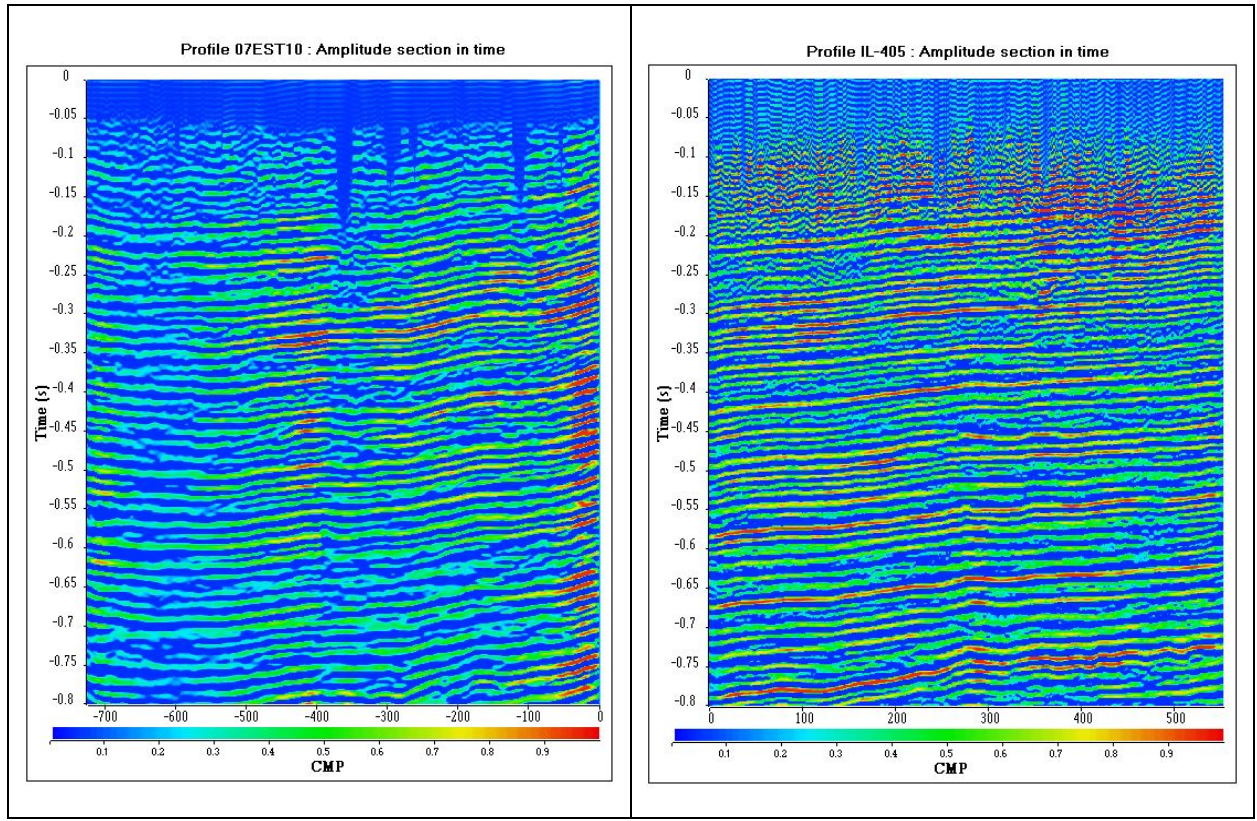

Figure 3

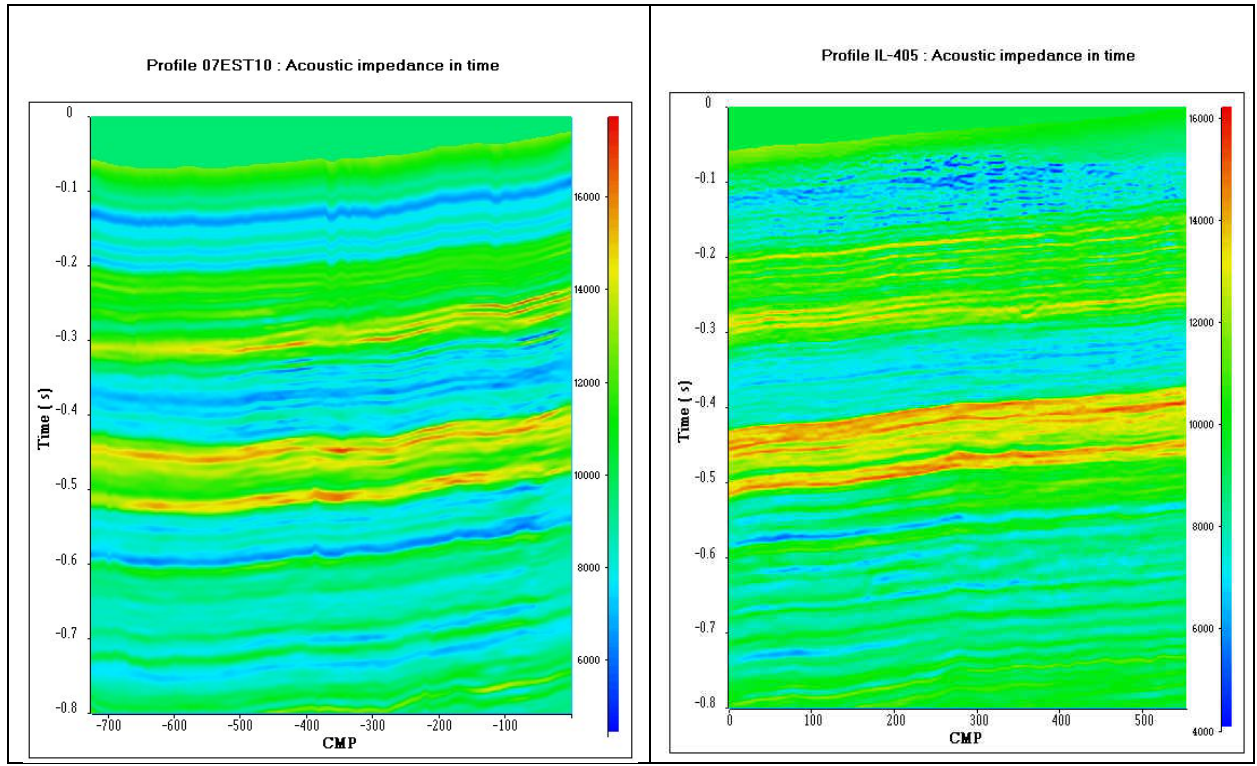

Figure 4

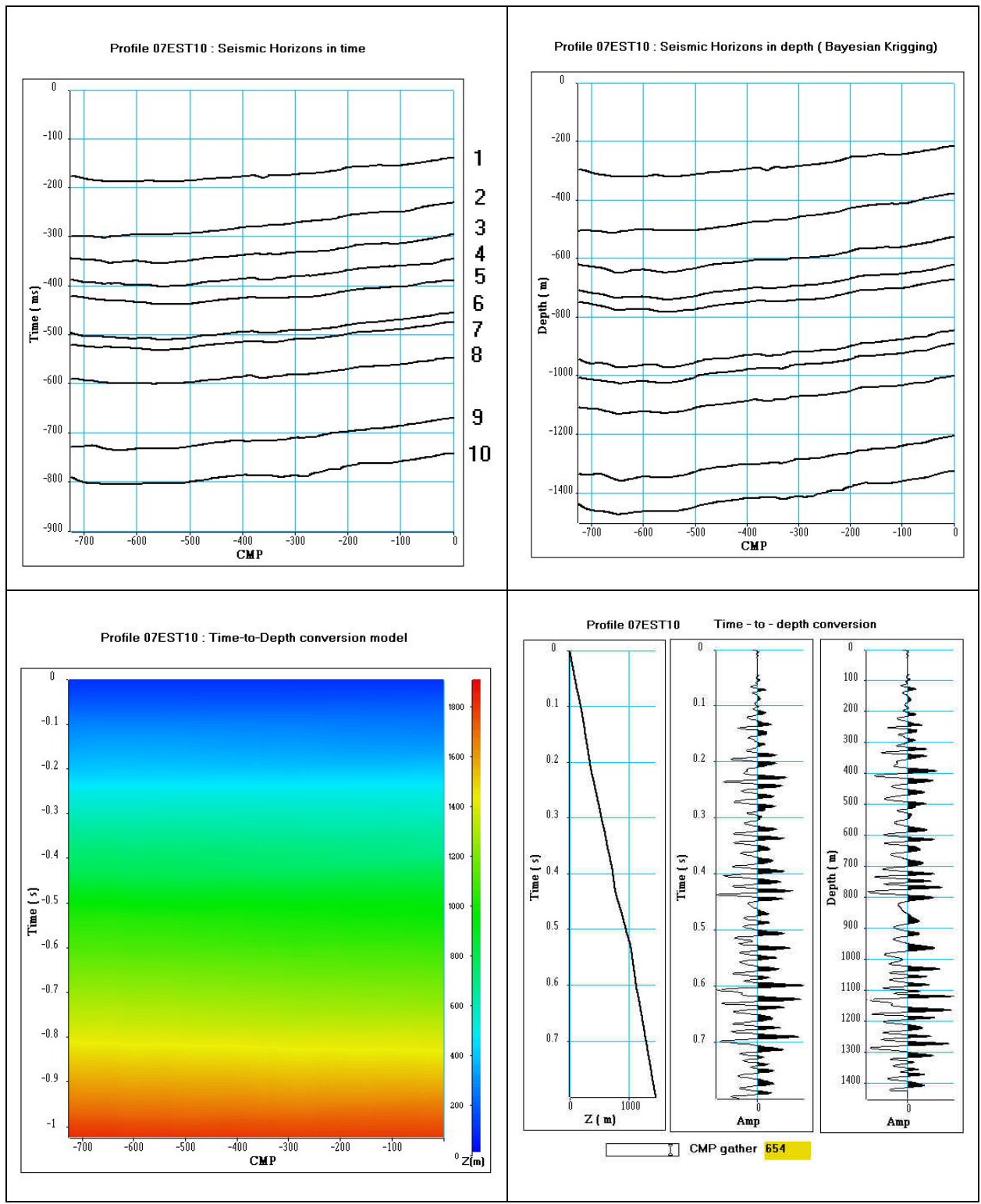

Figure 5

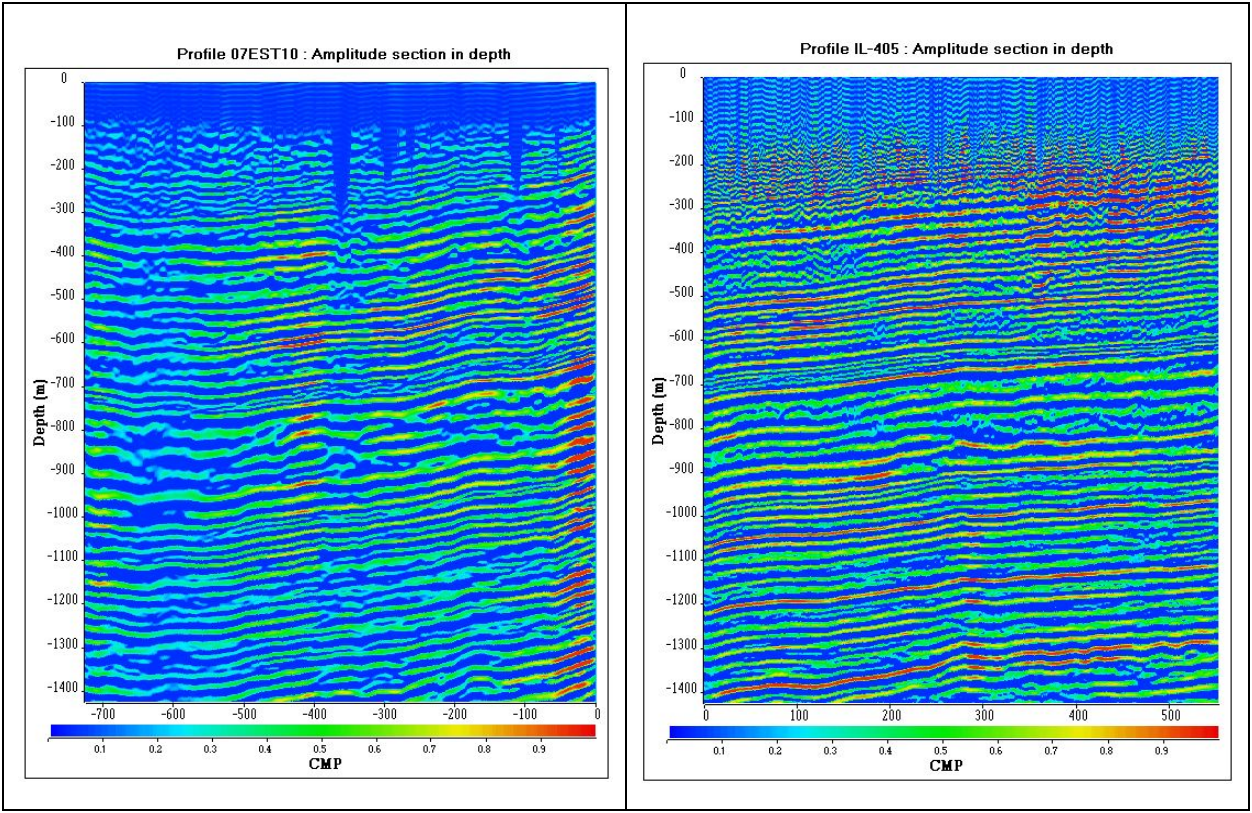

Figure 6

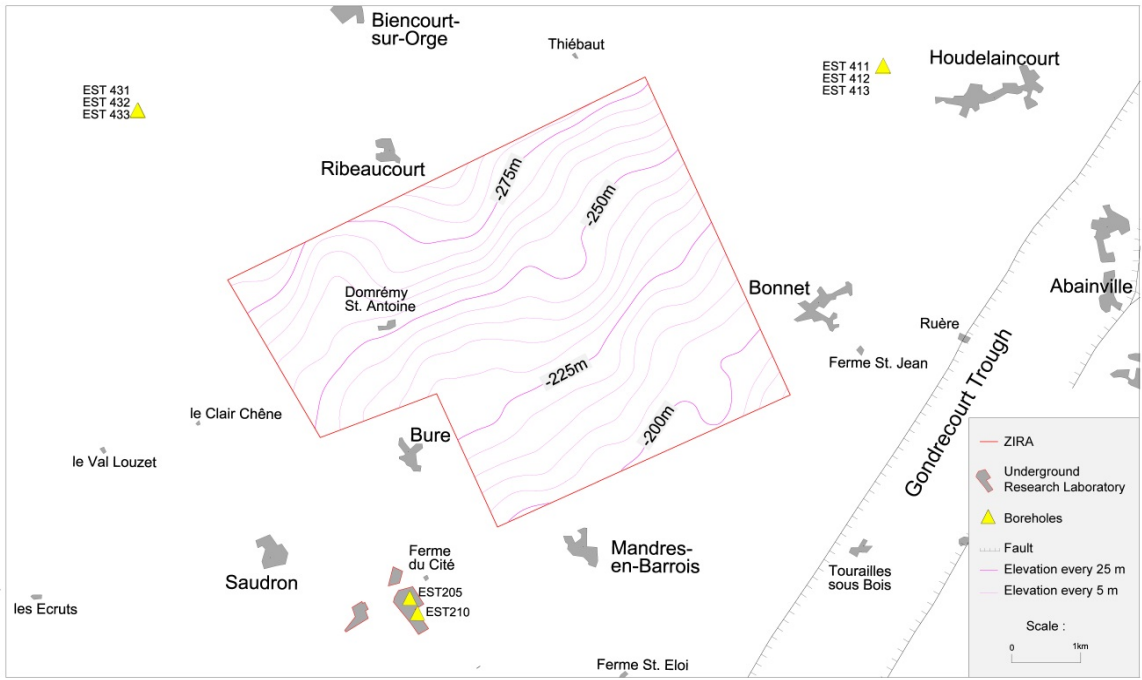

Figure 7

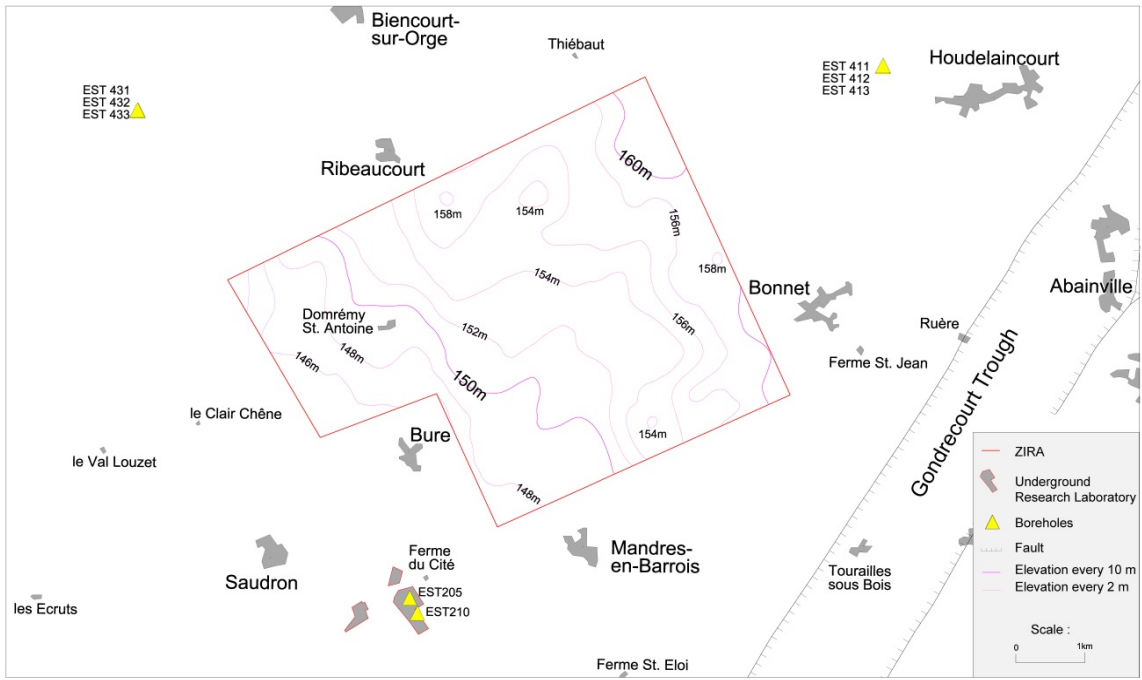

Figure 8

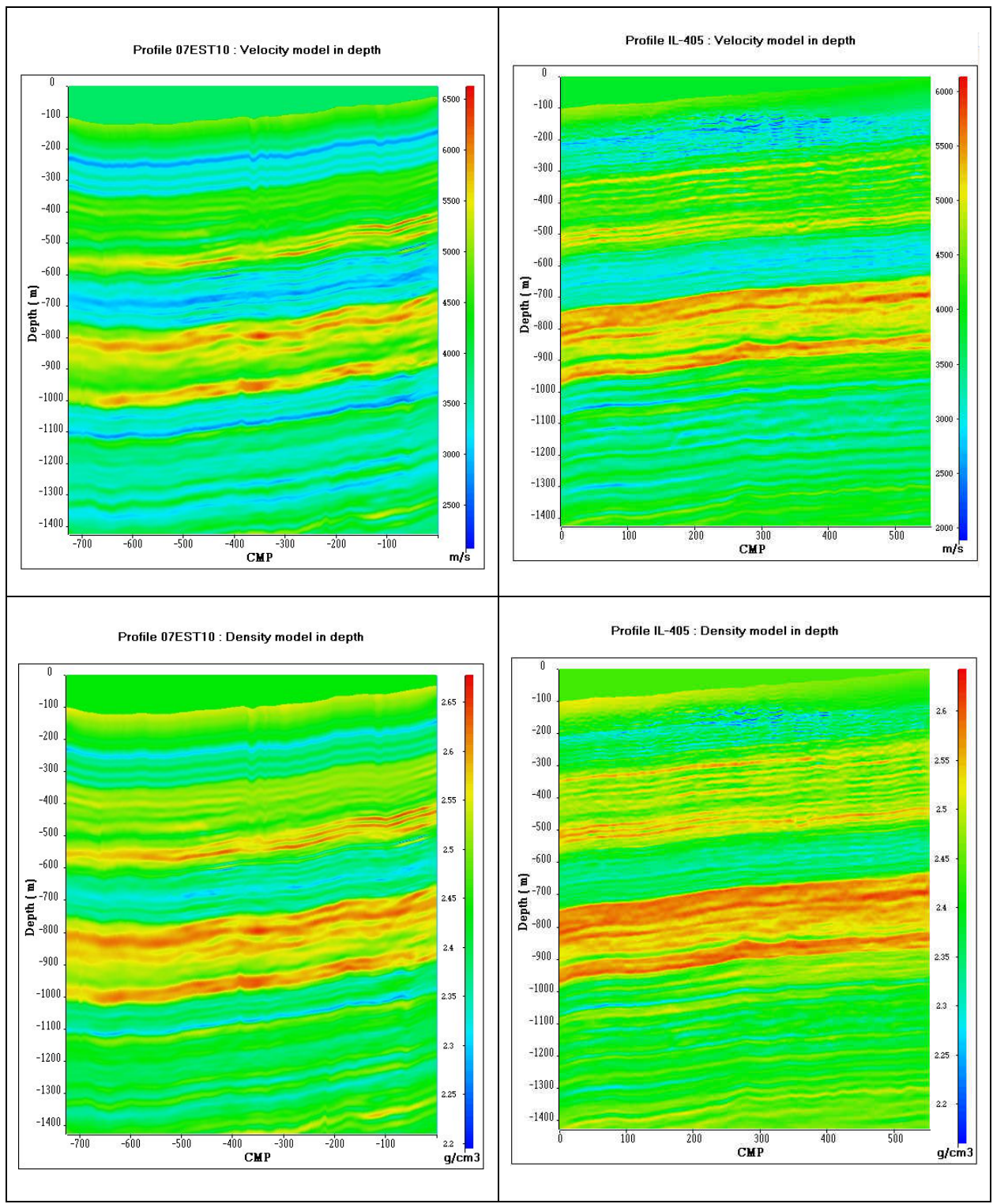

Figure 9

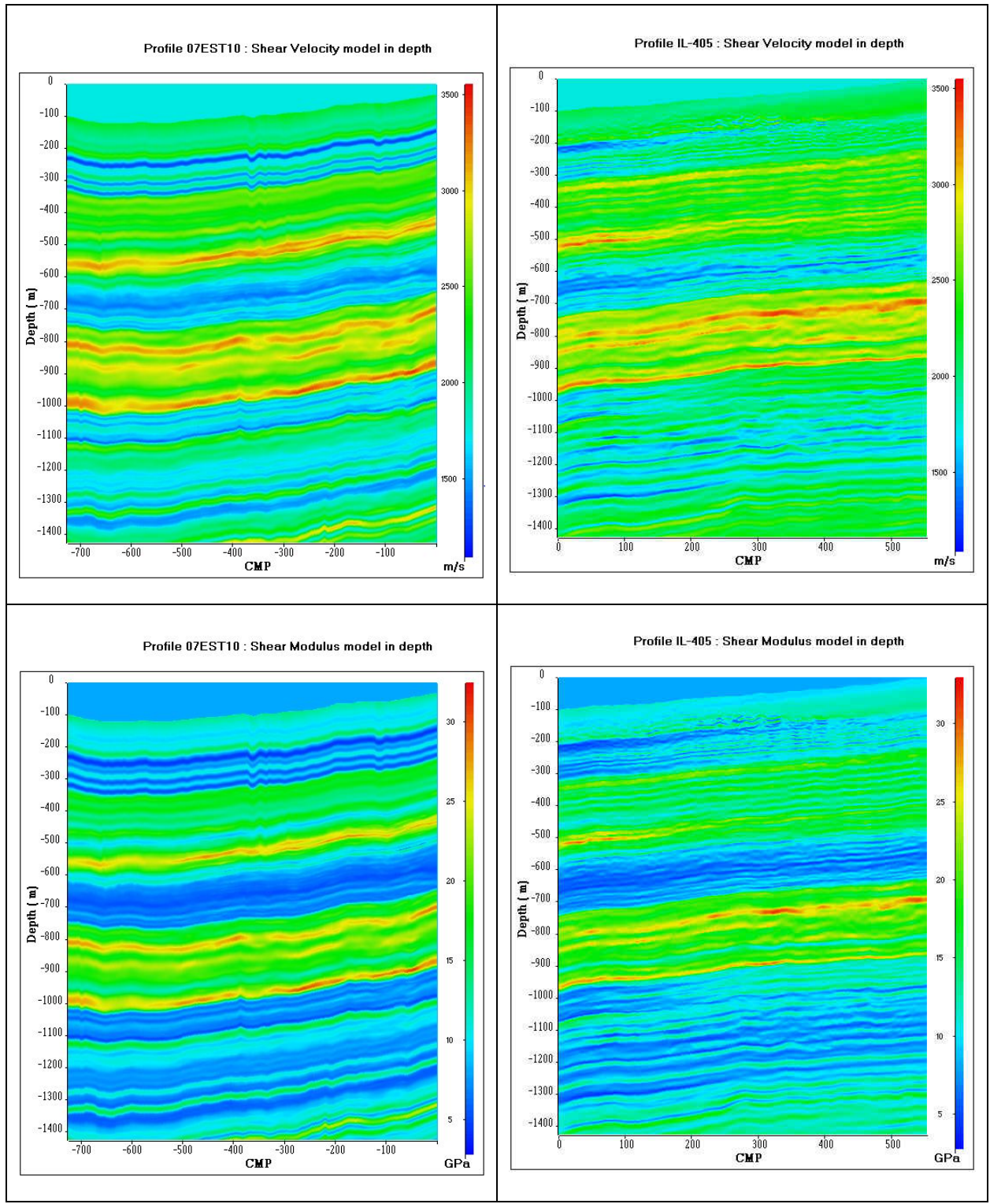

Figure 10

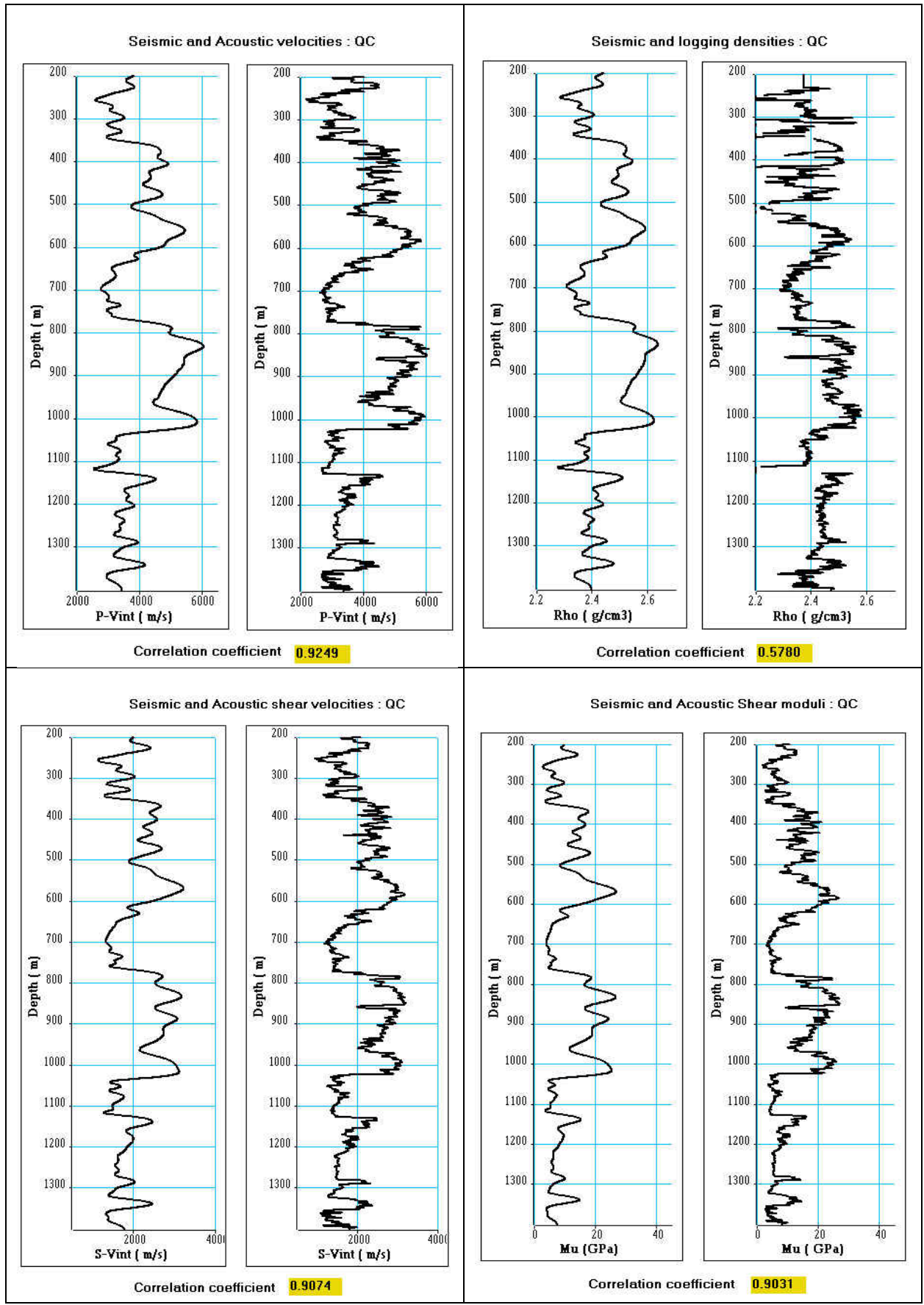

Figure 11

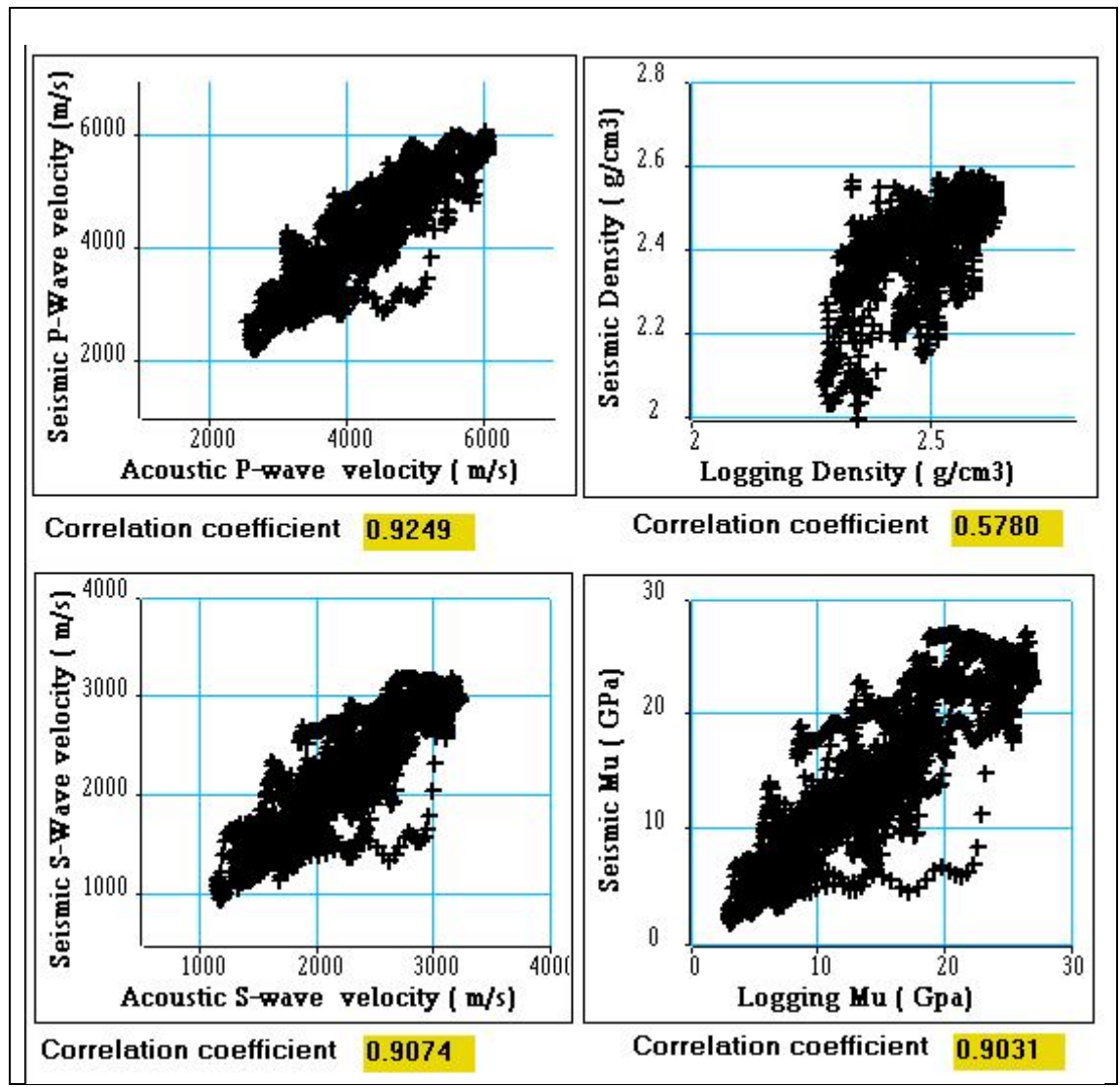

Figure 12

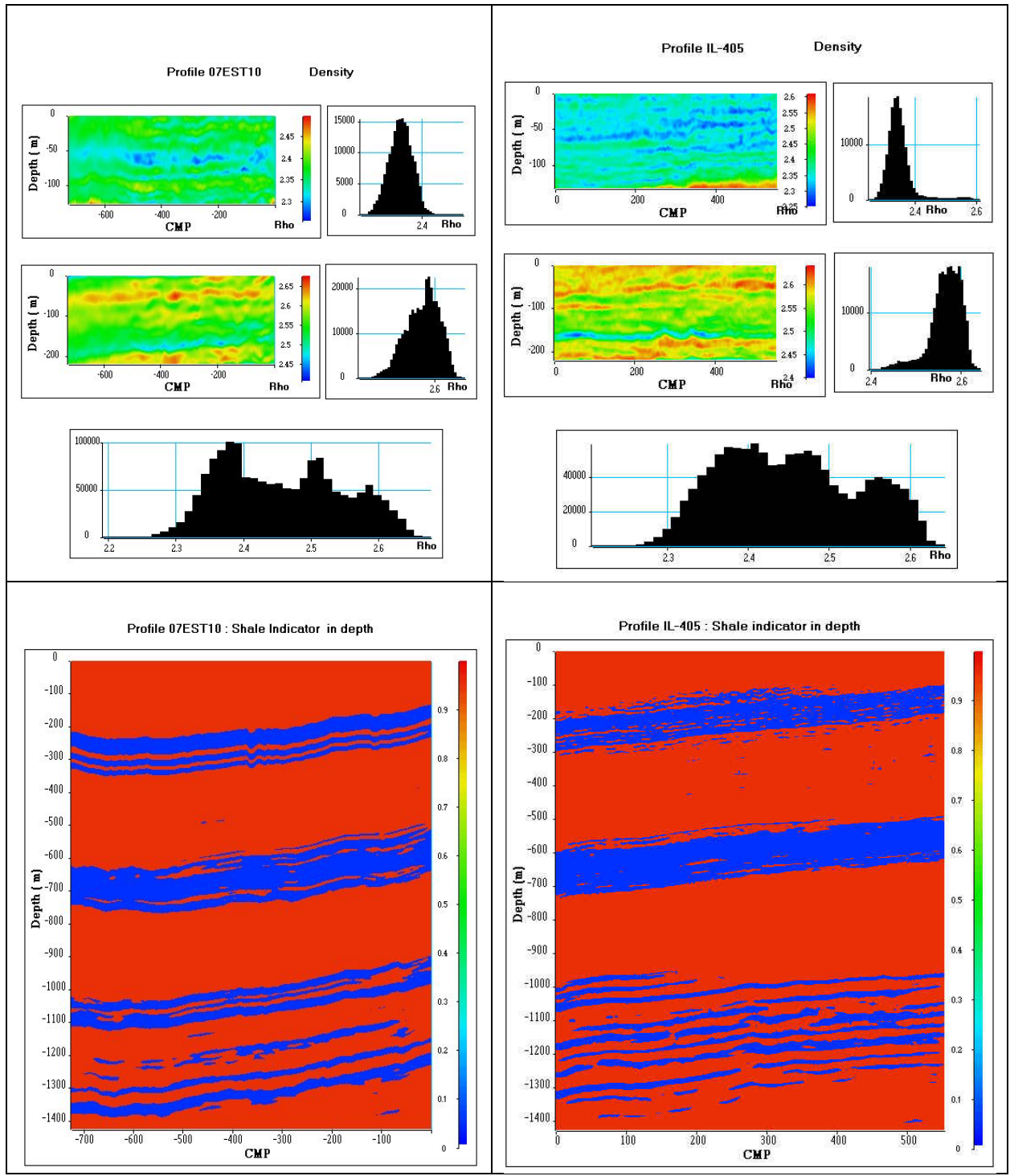

Figure 13

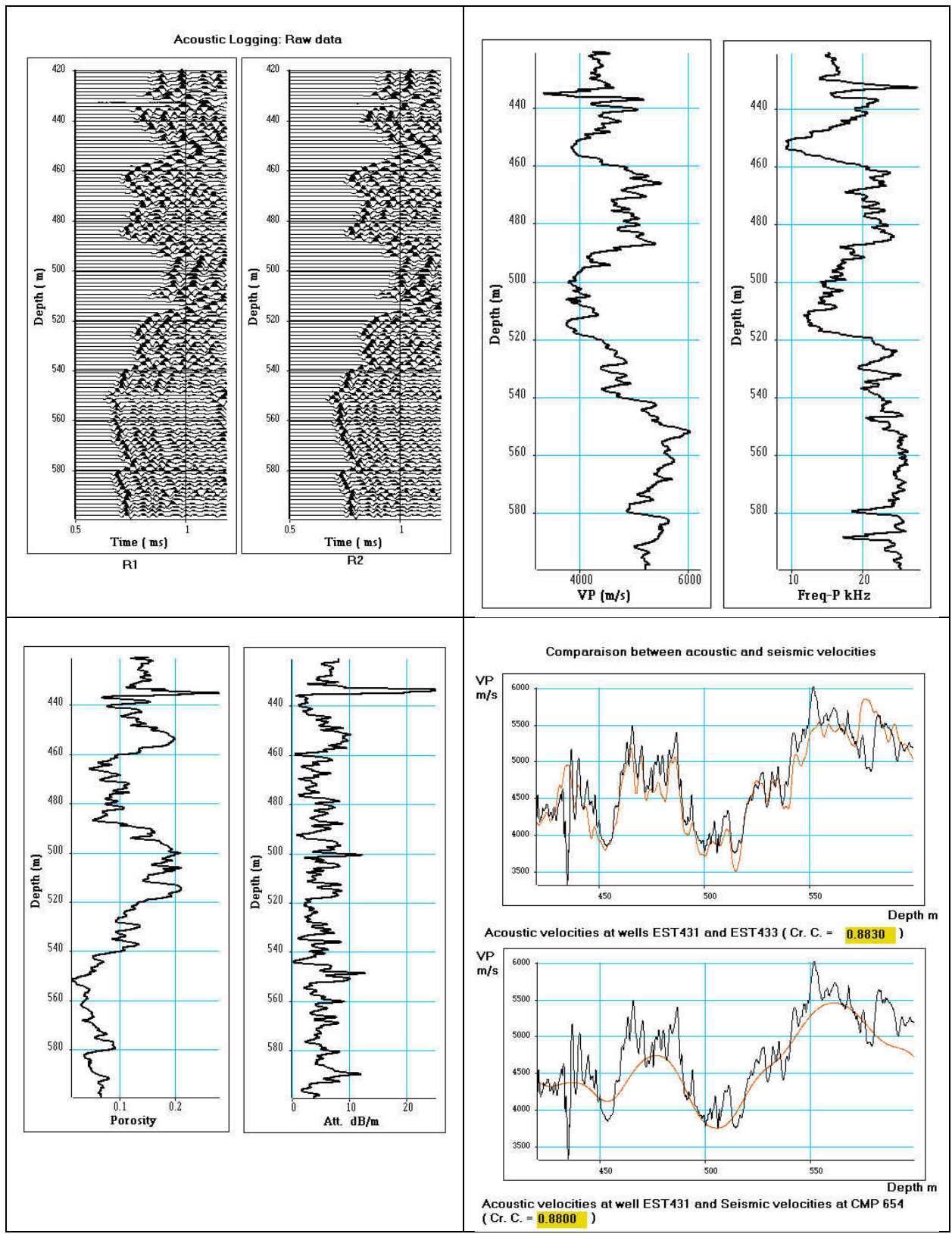

Figure 14

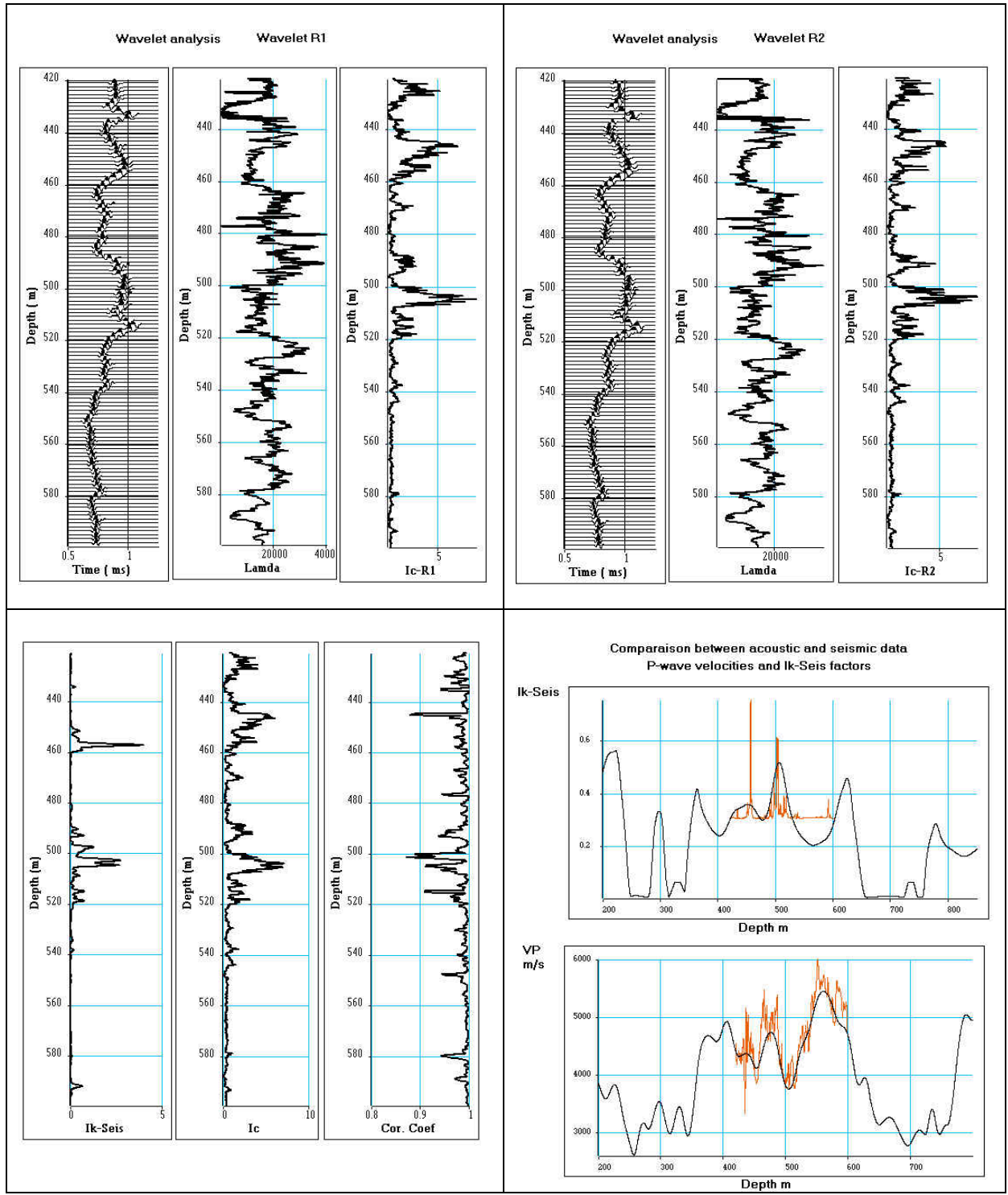

Figure 15

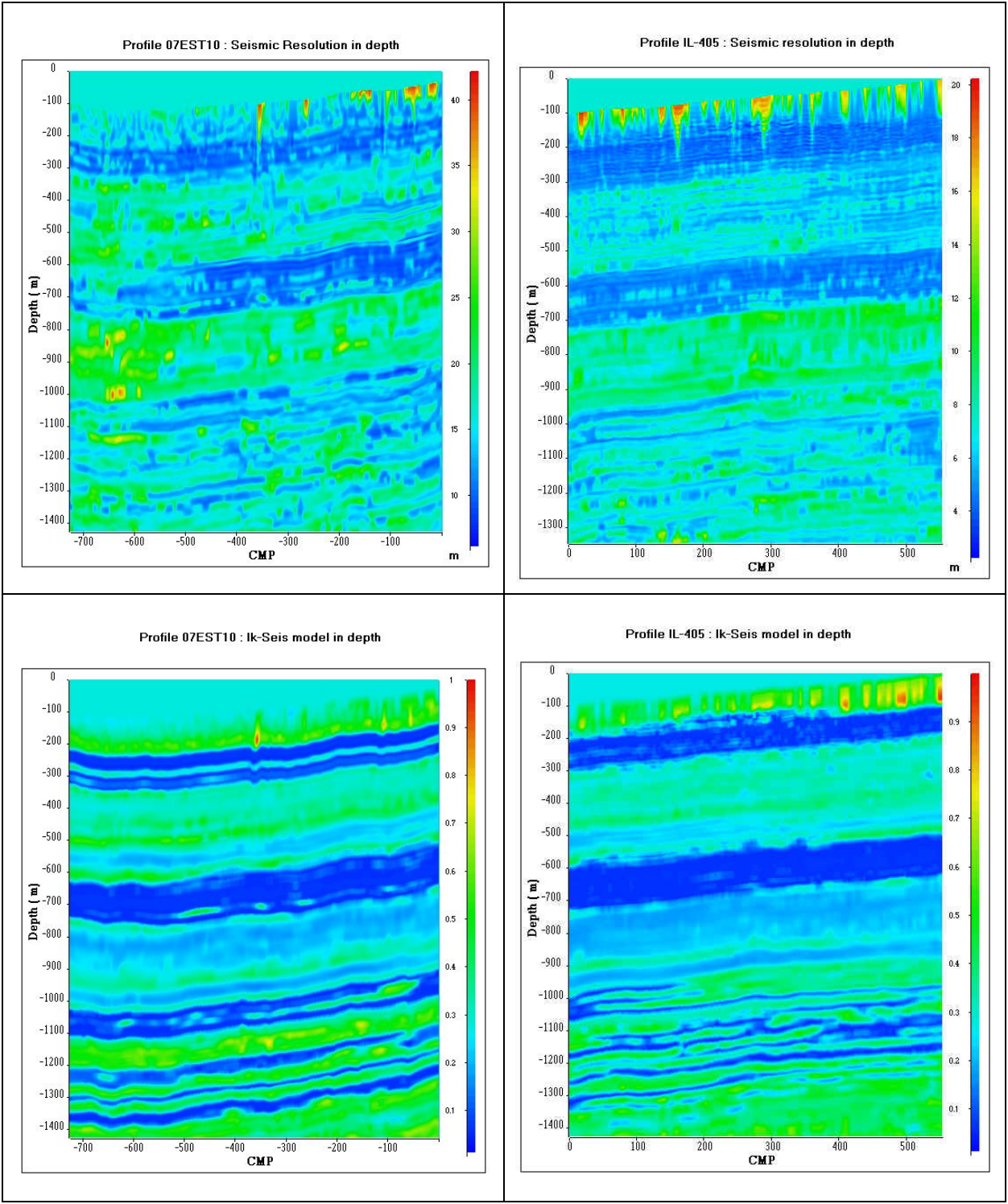

Figure 16

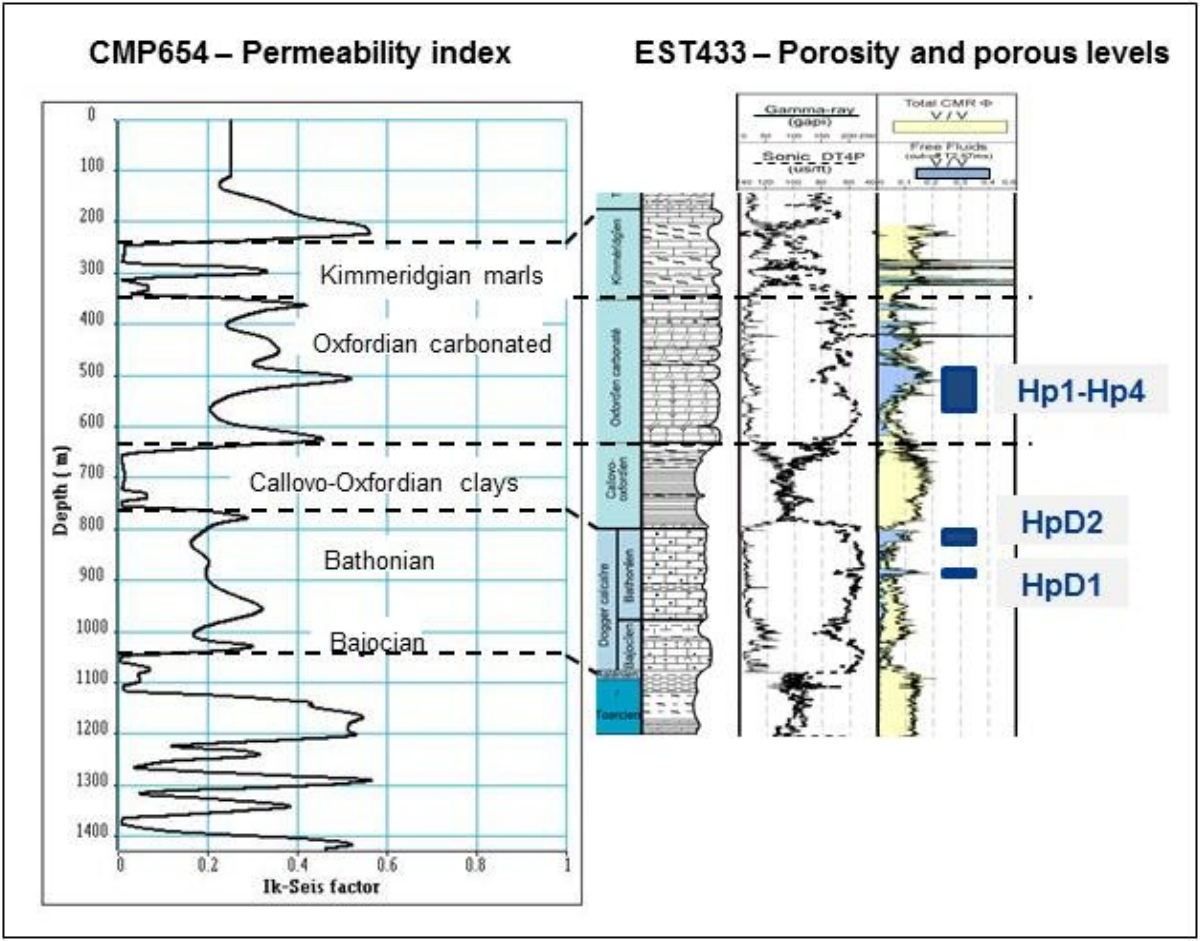

Figure 17